\documentclass{svjour3}                     
\smartqed  
\usepackage{graphicx}
 \journalname{Bulletin of Mathematical Biology}
\begin{document}

\title{Nonlinear waves in capillary electrophoresis
\thanks{supported by the NIH under grant  R01EB007596.}
}


\author{Sandip Ghosal       \and   Zhen Chen}


\institute{S. Ghosal \at
              Northwestern University, Dept. Mech. Eng., 2145 Sheridan Road, Evanston, IL 60208\\
              Tel.: +1 847 467 5990\\
              Fax: +1 847 491 3915\\
              \email{s-ghosal@northwestern.edu}           
           \and
           Z. Chen \at
           Northwestern University, Dept. Mech. Eng., 2145 Sheridan Road, Evanston, IL 60208\\
              Tel.: +1 847 491 3663\\
              Fax: +1 847 491 3915\\
              \email{zhenchen2011@u.northwestern.edu}    
}

\date{Received: date / Accepted: date}

\maketitle

\begin{abstract}
Electrophoretic separation of a mixture of chemical species is a fundamental technique of great usefulness in biology, health care and forensics. In capillary electrophoresis 
the sample migrates in a microcapillary in the presence of a background  electrolyte. When the 
ionic concentration of the sample is sufficiently high, the signal is known to 
exhibit features reminiscent of nonlinear waves including sharp concentration `shocks'.
 In this paper we consider a simplified model consisting of a single sample ion 
and a background electrolyte consisting of a single co-ion and 
a counterion in the absence of any processes that might change the ionization states of 
the constituents. If the ionic diffusivities are assumed to be the 
same for all constituents the concentration of sample ion is shown to obey a one dimensional advection 
diffusion equation with a concentration dependent advection velocity. 
If the analyte concentration is sufficiently low in a suitable non-dimensional sense, Burgers' equation 
is recovered, and thus, the time dependent problem is exactly solvable with arbitrary initial conditions. 
In the case of small diffusivity either a leading edge or trailing edge shock is formed depending on the 
electrophoretic mobility of the sample ion relative to the background ions. Analytical formulas 
are presented for the shape, width and migration velocity of the sample peak and it is shown that axial dispersion at 
long times may be characterized by an effective diffusivity that is exactly calculated. These results are consistent with known observations from physical and numerical simulation experiments.
\end{abstract}

\section{Introduction}
\begin{figure*}
\includegraphics[width=1.0\textwidth]{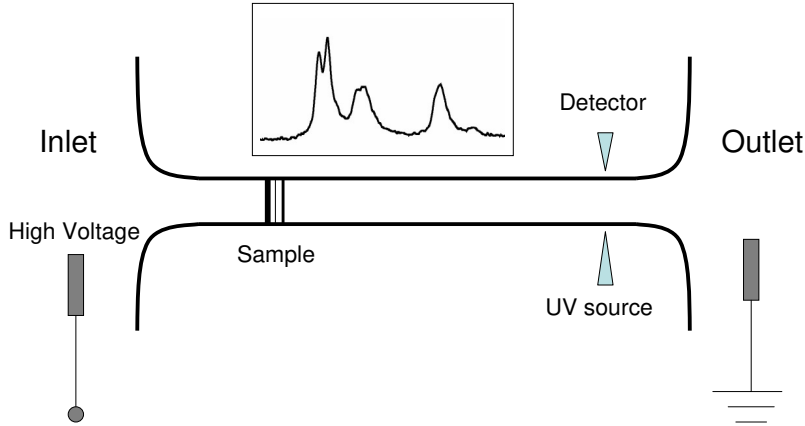}
\caption{A schematic sketch showing the CE setup and the electropherogram (inset) recorded by the detector.
The diagram is not to scale.}
\label{fig:0}      
\end{figure*}

Slab gel electrophoresis has long been a familiar technique in molecular biology for sorting biomolecules 
according to their size and charge. For example, the Sanger method of sequencing DNA and 
RNA uses gel electrophoresis for sorting the products of the chain termination reaction. The 
SDS-PAGE technique used for sorting proteins by size uses electrophoresis in a polyacrylamide 
gel to separate the surfactant coated amino acid chains resulting from the denaturing of the proteins.
The technique of Capillary Electrophoresis (CE) was introduced about thirty years ago and 
has largely replaced gel electrophoresis in many modern applications. In CE, the sample migrates 
through a microcapillary (about 50 $\mu$m in diameter and tens of centimeters long) instead of 
a slab of porous gel. The micro-capillary is connected at the inlet and at the outlet to reservoirs, 
as shown in the schematic sketch in Figure~\ref{fig:0} and the entire system is filled with an 
electrolytic buffer~\cite{czebook1}. A large electrical voltage (tens of kilovolts) is applied 
with electrodes between the inlet and outlet reservoirs. 
The sample separates into distinct zones based on the migration speeds of individual components in the applied 
electric field.  A detector placed near the outlet records a series of peaks (an electropherogram) 
that may be used for analyzing the sample. Often 
(but not always) there is an electroosmotic flow~\cite{probstein} present in the capillary that causes both positive and 
negative ions to be swept past a single detector placed near the outlet. For linear polyelectrolytes (e.g. DNA)
the mobility is insensitive to the polymer length. In such cases a sieving medium such as a polyacrylamide 
gel may be loaded into the micro-capillary. 

CE has many advantages over the traditional slab gel electrophoresis 
method, one of the most important ones being ease of parallelization. Since many CE channels can be etched on a 
single substrate, multiple separations can be run 
simultaneously. Such scalability  is critical to applications such as genome sequencing, and indeed 
CE is the method of choice in most such applications.
Though CE is generally considered superior to the more traditional gel electrophoresis methods one of its limitations is the high demands placed on the sensitivity 
of the photo detector. The detector, which consists of a UV or laser source coupled to a photo cell
must be sensitive to light attenuation over very short optical path lengths (the capillary diameter). 
Therefore, in order to maximize the likelihood of detection, it is preferable to use high sample concentrations.
This requirement however conflicts with the other requirement of CE which is to minimize 
peak broadening or dispersion in order to improve resolution and signal strength~\cite{ghosal_annrev06}.
The conflict is caused by a phenomenon known as ``electromigration dispersion'' or the ``sample 
overloading effect''. The underlying physical mechanism is  well understood and may be 
roughly explained as follows: when the concentration of sample ions is non-negligible compared 
to that of the background electrolytes 
the local electrical conductivity ($\sigma$)  is altered in the vicinity of the peak. On the other hand, since the current ($J$) 
is constant, the electric field ($E$) must change locally since by Ohm's law $J = \sigma E$. This 
varying electric field alters the effective migration speed of the sample (or analyte) ion which in turn alters 
its concentration distribution thereby giving rise to a nonlinear transport problem that must be solved in a self 
consistent manner. 
Electromigration dispersion causes highly asymmetric concentration profiles, high rates 
of dispersion and shock like structures that are reminiscent of nonlinear waves seen in many other physical systems. 
These effects have been widely reported in the literature on electrophoresis (see e.g.~\cite{bouskova_elph04}).

Simple one dimensional mathematical models of electromigration dispersion have been constructed~\cite{weber_die_1910,mi_ev_ve_79a,mikkers_ac99,math_th_elph_bk}
by invoking the assumption of vanishing  diffusivity  of all species and  the requirement of local charge neutrality, which for a three ion system yields a single nonlinear hyperbolic equation. 
Solutions describe the  observed steepening of an intitially smooth profile leading to subsequent shock formation. Two recent reviews~\cite{gas09,thormann09} provide more extensive 
references to one dimensional models and the behavior of their solutions as evident from numerical simulations. These models assume that only strong electrolytes are involved 
so that the valence states of the constituent ions are fixed. This assumption is inadequate when ampholytes (such as proteins) are present. The separation of ampholytes is usually performed 
by exploiting  two other modes of electrophoresis: Isotachophoresis and Isoelectric focusing. The mode of electrophoresis described in the introduction and discussed 
in this paper by contrast is called ``Zone electrophoresis''.  From a mathematical perspective the various modes of electrophoresis correspond to different kinds of initial 
and boundary conditions of the same governing equations. An excellent discussion of the mathematical problem involved in the various modes of electrophoresis may be found in 
Chapter~3 of the book 
by Fife~\cite{fife_dynamics_1988} and the practical aspects are discussed in a number of text books~\cite{czebook1,czebook2}. The essential technique for reducing the 
complexity of the transport equations in the presence of ampholytes was developed by Saville and Palusinski~\cite{saville_theory_1986} using the well founded 
assumption that the time scale associated with ionization phenomena is very short compared to the time scale of diffusive or electrophoretic transport. 
Thus, local equilibrium may be assumed to exist between the ampholyte and its dissociation products so that the concentrations of the various charged states
can be algebraically related to the concentration of the ampholyte in the neutral state. A historically contentious issue and an area of some mathematical 
interest concerns the assumption of local charge neutrality mentioned earlier. Local charge neutrality results when the small parameter (in some appropriate 
dimensionless formulation) multiplying the Laplacian of the potential in Poisson's equation vanishes. This assumption, first invoked by Planck~\cite{planck}, 
results in a singular perturbation problem  leading to boundary layers and associated interesting physics in various areas including semiconductor junctions and 
in selectively permeable membranes~\cite{rubinstein_book}. In the context of electrophoresis, the singular nature of the local charge neutrality assumption is manifested 
at the sharp boundaries of isotachophoretic shocks where the shock is regularized by finiteness of the small parameter. 
This aspect has been studied by Fife~\cite{fife_electrophoretic_1988} who also presented some results relating to the regularity of solutions. A proof of the global 
existence of solutions for the coupled species and electric field equations in one dimension in the absence of ampholytes have been presented by  
Avrin~\cite{avrin_global_1988}. In the present paper, we consider the one dimensional model of electromigration dispersion in the absence of ampholytes, however,  we replace the assumption 
of zero ionic diffusivities with one of equal diffusivity of the ionic constituents. We show that this simple modification leads to a  nonlinear 
advection-diffusion equation which reduces to Burgers' equation for the analyte concentration if this concentration is not too large. The wealth of information 
that is available about the Burgers' equation can thereby be exploited for understanding electromigration dispersion.  

\section{Fundamental equations of transport}

We consider a system of $N$ ionic species and denote by $c_{i} (x,t)$ the number of ions of species $i$ ($i=1,2,\ldots,N$) present per unit 
volume. Here $x$ is the direction 
along the capillary and $t$ is time. The system is presumed one dimensional, that is, the dependent variables are assumed 
not to vary over the cross-section of the capillary. In particular, this implies, electroosmotic flow, if present, is of the ``plug flow''
type. Cross stream variations can and do arise in CE and have been widely studied but it is not 
central to the present discussion. The concentrations obey the equations 
\begin{equation} 
\frac{\partial c_{i}}{\partial t} + \frac{\partial q_{i}}{\partial x} = 0 
\label{conserve_i}
\end{equation} 
where $q_{i}$ is the flux (amount passing  per unit area of cross section per unit time) of species $i$. In writing Eq.~(\ref{conserve_i}) 
we have presumed that there are no sources or sinks for the ions. In particular, ionization and recombination phenomena are ignored, which is justified if only strong electrolytes are involved.
The fluxes $q_{i}$ are related 
to the concentration fields through the Nerst-Planck model
\begin{equation} 
q_{i} (x,t) = - D_{i} \frac{\partial c_{i}}{\partial x} + \mu_{i} c_{i} (x,t) E (x,t),
\label{Nerst-Planck}
\end{equation}
where $E(x,t)$ is the electric field (directed along the capillary), $\mu_{i}$ is the electrophoretic mobility
and $D_{i}$ is the diffusivity of the `$i$th' species.
 In addition to the electrophoretic mobility, which is the 
velocity acquired by the ion per unit of applied electric field, we would also refer to the mobility 
($u_{i}$) which is the velocity per unit applied force. The two are related\footnote{this relation is not valid 
for macro-ions or colloidal particles owing to Debye layer effects. The relation is valid provided the ionic 
size is much smaller than the Debye length} to each other as $\mu_{i} = z_{i} e u_{i}$ 
and related to the diffusivity through the Stokes-Einstein relation~\cite{russel_saville_schowalter}
 $D_{i}/u_{i} = k_{B} T$, where $k_{B}$ 
is Boltzmann's constant, $T$ is the absolute temperature, $e$ denotes the proton charge 
and $z_{i}$ is the signed valence of the $i$th species which can be positive, negative or zero. 
When fluid flow with velocity $V$ is present, a term $V c_{i}$ is added to the right hand side of 
Eq.~(\ref{Nerst-Planck}). However, if the flow is an electroosmotic plug flow, this term may be absorbed in the second term 
by simply redefining $\mu_{i}$ as the ``total'' instead of  the ``electrophoretic" mobility. 
If the capillary is filled with a polymer network then the mobility and diffusivity introduced above 
should be interpreted as the ``effective'' values of these quantities in the porous medium and the dependent 
variables must be regarded as averages of the corresponding physical quantity over the cross-section 
of the capillary. 

The system of equations Eq.~(\ref{conserve_i}) - (\ref{Nerst-Planck}) is closed by Poisson's 
equation of electrostatics, which in SI units is written as 
\begin{equation}
\epsilon \frac{\partial E}{\partial x} = \rho_{e} (x,t) = \sum_{i=1}^{N} e z_{i} c_{i} (x,t).
\label{poisson}
\end{equation}
Here  $\rho_{e} (x,t)$ is the charge density and $\epsilon$ is the permittivity  of the medium.  Time variations in electrophoresis 
are sufficiently slow that any errors resulting from replacement of the complete Maxwell's equations of electromagnetism by 
Poisson's equation, Eq.~(\ref{poisson}),  is negligible. Eq.~(\ref{conserve_i}) -- (\ref{poisson}) provide a closed 
system of equations for calculation of the time evolution of the functions $c_{i}(x,t)$ if the initial conditions are known and if 
appropriate boundary conditions are specified.

\section{Electroneutrality and the Kohlrausch function}
If `$a$' denotes a characteristic length scale over which the concentrations may be considered to vary 
(this may be taken, for example, as the peak width), then it may be shown
that the ratio of the term on the left hand side of Eq.~(\ref{poisson}) to that on the right is of the order of 
$(\lambda_{D} / a )^{2}$, where $\lambda_{D}$ is the Debye length in the buffer. Since in CE, $a$ is of 
the order of microns while $\lambda_{D}$ is in the nanometer range, 
$(\lambda_{D} / a )^{2}$ is extremely small. As a result, Eq.~(\ref{poisson}) may be replaced by the 
electroneutrality condition:
\begin{equation}
\rho_{e} (x,t) = \sum_{i=1}^{N} e z_{i} c_{i} (x,t) = 0,
\label{electroneutral}
\end{equation}
first invoked by Planck~\cite{planck} in the context of the problem of the liquid junction potential. This condition caused some 
confusion in the early history of the subject, since it appeared that in the absence of the charge density, no 
electric field, or in our problem, perturbation to the applied field is possible. The controversy was resolved~\cite{Hickman} when it was realized that Eq.~(\ref{electroneutral}) should be interpreted in an asymptotic 
rather than in an exact sense.

The disappearance of the left hand side of Eq.~(\ref{poisson}) may at first sight appear to be  problematic because 
we have now lost our equation for determining $E$. This is not in fact a difficulty as $E$ can be determined 
directly from Ohm's law and the condition of the constancy of current:
\begin{equation}
J_{0} = J = \sigma E 
\label{ohm}
\end{equation}
where the electrical conductivity is\footnote{the dependance of conductivity on concentrations is actually nonlinear 
at high concentrations~\cite{robinson_stokes_book}. However we ignore this effect as it does not fundamentally
alter our analysis except in replacing one nonlinear dependence of ion velocity on concentration by another}
\begin{equation} 
\sigma = \sum_{i=1}^{N}  \mu_{i} z_{i} e c_{i} =  e^{2} \sum_{i=1}^{N} z_{i}^{2}   u_{i}  c_{i}
\label{conductivity}
\end{equation} 
and $J_{0}$ is a constant. The first of the two equalities in Eq~(\ref{ohm}) is a consequence of 
charge neutrality but the second requires certain additional assumptions.
To see this, substitute $q_{i}$ from Eq.~(\ref{Nerst-Planck}) in the expression for the current
\begin{equation} 
J = \sum_{i=1}^{N} q_{i} z_{i} e = - \frac{\partial}{\partial x} \left[ \sum_{i=1}^{N} e z_{i} c_{i} D_{i} \right] 
+ E \sum_{i=1}^{N}  \mu_{i} z_{i} e c_{i}.
\label{expressionJ}
\end{equation} 
If the diffusive contributions are now neglected, then  the second part of Eq.~(\ref{ohm}), 
$J = \sigma E$ or  Ohm's law  follows. 

The constraints imposed by Eq.~(\ref{electroneutral}) and (\ref{ohm}) have the effect of replacing 
two of the differential equations governing the transport by algebraic relations. 
 It was pointed out by Kohlrausch~\cite{kohlrausch} that a further reduction of the governing equations is possible if 
the diffusion terms are neglected in the ion transport equations.
To see this we define the Kohlrausch function: 
\begin{equation} 
K(x,t) = \sum_{i=1}^{N} \frac{c_{i}}{u_{i}}.
\label{kohlrauch}
\end{equation}
If we now take the derivative of both sides of this relation with respect to time and use Eq.~(\ref{conserve_i}),
(\ref{Nerst-Planck}) and (\ref{electroneutral}) we get 
\begin{equation} 
\frac{\partial K}{\partial t} =  \frac{\partial^{2}}{\partial x^{2}} \left( \sum_{i=1}^{N} \frac{D_i}{u_i}c_{i} \right).
\label{Kohlrausch_eqn}
\end{equation} 
If the right hand side is neglected we get 
\begin{equation}
K(x,t) = K(x,0),
\label{Kisconstant}
\end{equation} 
and this provides an algebraic relation to replace one more  differential 
equation for transport. 

In contrast to the electro-neutrality condition which is respected to high accuracy, 
the neglect of the diffusion terms in arriving at either Ohm's law or the Kohlrausch  
relation is questionable in applications related to CE.
The ratio of the diffusive term to the other terms is measured by the inverse of the Peclet number,
${\rm  Pe} = v a / D$ in terms of the characteristic peak width $a$ and velocity $v$.
While it is true that the Peclet number  is generally large in the applications we are concerned with here, 
the  transit times  $L/v$ across the length ($L$) of the capillary is also very long. Thus, the ratio 
of the diffusive  broadening ($\sqrt{2 D L / v}$) to a characteristic peak width ($a$)  is measured by the
square root of $(L/a) {\rm Pe}^{-1}$ which is generally not small even though ${\rm Pe} \sim 10 - 100$. 
Furthermore, when the profile steepens, as it does in electromigration dispersion, 
the second derivative term multiplying $D_{i}$ in the transport equations can become 
very large so that the diffusive term once again has a finite effect no matter how small the diffusivity is.
Nevertheless, the Kohlrausch relations have been fruitfully applied for certain special situations 
such as in calculating fluxes across interfaces separating two homogeneous regions. In 
the next section we show how the ideas of Kohlrausch and Ohm's law can be generalized to 
treat the case of nonzero diffusivities.

\section{Reduction to a one equation model}
Let us now consider an idealized model of electrophoresis where only three species 
are involved: a sample or analyte ion, a positive ion and a negative ion. In place of the suffix $i$ we will 
from now on use a suffix `$p$' to denote the positive ion, `$n$' to denote the negative ion 
and no suffix at all to indicate the sample. Thus, $z_{p}$ is the signed valence of the positive ion, 
$z_{n}$ that of the negative ion and $z$ that of the sample. Clearly $z_{p}$ is positive, $z_{n}$ 
is negative and $z$ could be of either sign (but not zero, since a neutral species does not 
electromigrate). We will further assume that all the species have the same mobility ($u$), 
and therefore the same diffusivity $D=uk_{B}T$, though their electrophoretric mobilities 
$\mu_{i} = e z_{i} u$ could be widely different. Many analytes of importance in electrophoresis are linear polyelectrolytes, and for these 
molecules the charge scales linearly with the mass $M$ while the diffusivity scales roughly
as $M^{-1/3}$. It might therefore be argued that diffusivities of ions vary over a relatively 
small range when compared to their electrophoretic mobilities, though it must 
be acknowledged that our primary motivation for introducing this assumption is the pragmatic one  of  wanting to
reduce the mathematical complexity of the problem.

\begin{figure*}
\includegraphics[width=1.0\textwidth]{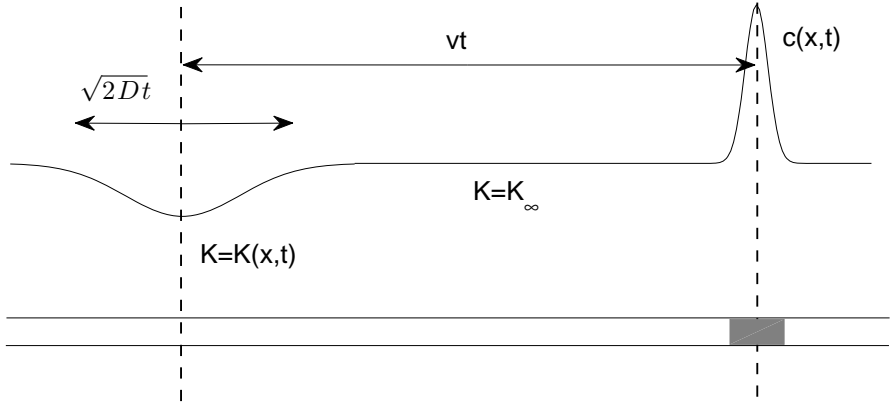}
\caption{Sketch illustrating the basis of the assumption that the Kohlrausch function 
may be assumed constant in the sample zone shortly after sample injection.}
\label{fig:1}      
\end{figure*}
It is readily seen that by virtue of our assumption $D_{i} = D$ and the 
electroneutrality condition Eq.~(\ref{electroneutral}), the first term of Eq.~(\ref{expressionJ})
vanishes so that Ohm's law, Eq~(\ref{ohm}) is recovered. Further, 
in view of our assumption of equal diffusivities, Eq.~(\ref{Kohlrausch_eqn}) reduces to the diffusion equation 
\begin{equation} 
\frac{\partial K}{\partial t} = D \frac{\partial^{2} K}{\partial x^{2}} 
\end{equation} 
which admits an analytical solution:
\begin{equation} 
K(x,t) = \frac{1}{\sqrt{4 \pi Dt }} \int_{-\infty}^{+\infty} K(y,0) \exp \left[ - \frac{(x-y)^{2}}{4 D t} \right]  \; dy.
\end{equation} 
If $K_{\infty}$ denotes the unperturbed value of $K$, then the perturbation in $K$, 
$\delta K = K - K_{\infty}$ spreads diffusively from the injection region so that its width,
$w \sim \sqrt{w_{0}^{2} + 2 D t}$, where $w_{0}$ is the width of the peak at injection.
The essential observation that allows us to extend Kohlrausch's ideas to the diffusive case 
is illustrated in Figure~\ref{fig:1} and it is the following:
even though $K(x,t)$ is in general a time dependent quantity, in the vicinity of the analyte peak
we may assume it to be a constant equal to its unperturbed value $K = K_{\infty}$. 
This is because the distance between the analyte peak and the injection 
zone ($X$) increases linearly with time $X \sim \mu E_{0}  t$ where $E_{0}$ is the applied field
so that at long times, $X \gg w$. The minimum length of time that needs to pass for this to happen 
is  ${\rm max} \, [ w_0/v, D / v^{2}]$ where $v = \mu E_{0}$ is the velocity of an isolated analyte ion 
in the back ground electrolyte and ${\rm max} \, [\quad]$ indicates the larger of the two numbers
within the brackets.  It is instructive to express  this in terms of the diffusion 
time $\tau_{d} = a^{2} / D$ the time it takes the analyte to diffuse over a characteristic distance. 
If we identify this characteristic distance with $w_0$, that is, $a = w_0$, then the condition 
may be written as $t / \tau_{d} \gg {\rm max} [ \, {\rm Pe}^{-1},{\rm Pe}^{-2} \, ]$ where ${\rm Pe} = v a /D$ 
is  the Peclet number introduced earlier. If we use the estimates 
$v \sim 1$ mm/s, $a = w_{0} \sim 10$ $\mu$m, $D \sim 10^{-5}$ cm$^{2}$/s which are fairly 
typical, we have ${\rm  Pe} \sim 10$ and $\tau_{d} \sim 0.1$ s. Thus, the condition $K = K_{\infty}$ 
in the vicinity of the analyte peak may be used with confidence for times $t \gg 10^{-2}$ s.
This is indeed very small in comparison to the time interval between injection and detection 
which is tens of minutes. In simple terms, $\delta K$ spreads so 
slowly that the analyte peak leaves it behind essentially in the time it takes to 
traverse a distance of the order of its own width. 

Setting the right hand side of Eq.~(\ref{kohlrauch}) to the constant value $K_{\infty}$ immediately 
yields a third algebraic relation in addition to Eq.~(\ref{electroneutral}) and Eq.~(\ref{ohm}) 
between the set of four dependent variables: $c_p$, $c_n$, $c$ and $E$. If we use these to 
express all of the remaining variables in terms of $c$ and substitute in the transport equation 
for $c$, we obtain, after some algebra that we omit, 
 a single nonlinear partial differential equation describing the evolution 
of the analyte concentration $c$. This equation may be conveniently expressed in terms 
of the dimensionless quantity $\phi = c / c_{n}^{(\infty)}$ where $c_{n}^{(\infty)}$ is the 
concentration of the negative ions in the background electrolyte:
\begin{equation} 
\frac{\partial \phi }{\partial t} + v \frac{\partial }{\partial x} \left(  \frac{\phi}{1 -  \alpha \phi} \right) = D 
\frac{\partial^{2} \phi}{\partial x^{2}}.
\label{eq4phi}
\end{equation} 
The dimensionless parameter $\alpha$ is defined as 
\begin{equation} 
\alpha = \frac{(z - z_{n})(z - z_{p})}{z_{n} (z_{p}-z_{n})} 
= \frac{(\mu - \mu_{n})(\mu - \mu_{p})}{\mu_{n} (\mu_{p}-\mu_{n})}.
\end{equation}
Since $\mu_{p} > 0$ and $\mu_{n} < 0$, $\alpha$ is positive if $\mu$ is in the interval 
$( \mu_{n}, \mu_{p})$  and is negative otherwise. The parameter $\alpha$ is essentially 
a dimensionless form of the ``velocity slope parameter'' that has been used~\cite{gebauer_bocek_ac97} to characterize 
electromigration dispersion, since $\alpha v = V^{\prime}(0)$ where $V(\phi) = v/(1-\alpha \phi)$ 
is the local value of the migrational velocity of an analyte ion within the sample zone.

The concentrations of the negative and positive 
ions may be readily expressed in terms of $\phi$ using the charge neutrality condition, Eq~(\ref{electroneutral}) 
and the defining relation for $K(x,t)$, Eq~(\ref{kohlrauch}) :
\begin{equation} 
c_n = c_{n}^{(\infty)} \left[ 1 + \frac{z - z_p}{z_p - z_n}  \phi \right] + \frac{\mu_p}{z_p-z_n} \, \delta K,
\label{cn}
\end{equation} 
\begin{equation} 
c_p = c_{n}^{(\infty)} \left[ - \frac{z_n}{z_p} - \frac{z - z_n}{z_p - z_n}  \phi \right] - \frac{\mu_n}{z_p-z_n} \, \delta K.
\label{cp}
\end{equation} 
It is seen from the above equations that once the analyte has moved out of the injection zone 
($\phi \rightarrow 0$) the background ions do not immediately return to their equilibrium 
concentrations. A fluctuation is created in the concentration of the background ions which 
does not propagate, but decays in place on a diffusive time scale. If the capillary carries an 
electroosmotic flow and if the sensor is sensitive to the background ions  (e.g. a 
UV absorbance detector) these fluctuations may be detected with the analyte peaks 
complicating the interpretation of the signal. The existence of these anomalous peaks are well
known~\cite{desiderio_JCA97,bullock_ac95,poppe_ac92} and are variously referred to as ``water peaks'', ``system peaks'' or ``vacancy peaks''.

When $\alpha$ is positive, the evolution equation for $\phi$, Eq~(\ref{eq4phi}) has a 
singularity at $\phi = 1/\alpha$ which needs to be understood. Clearly, this singularity 
arises because of the vanishing of the conductivity $\sigma$ in Ohm's law Eq~(\ref{ohm}).
Since $\mu_{i} = e z_{i} u$, the vanishing of the conductivity when $\phi \rightarrow 1 /\alpha$ implies that
one or both of the background ion concentrations must have become negative. This, of course 
is physically impossible and would imply that at least one of our two assumptions, namely 
charge neutrality or constancy of the Kohlrausch function would have to be violated before 
this singularity is reached. In fact, the highest analyte concentration to which Eq~(\ref{eq4phi}) 
may still be used can be calculated by requiring that both the conditions $c_n > 0$ and 
$c_p > 0$ be respected where $c_n$ and $c_p$ are given by Eq.~(\ref{cn}) and Eq.~(\ref{cp})
with $\delta K =0$.
These inequalities may be reduced after some algebra to the requirement $\phi < \phi_{c}$ 
where $\phi_{c} = (z_p - z_n)/(z_p - z)$ when $z < 0$ and $\phi_{c} = - [ z_n (z_p - z_n) / [ z_p (z - z_n) ]$
when $z > 0$. As a practical example, if $z = \pm 2 z_p$ and $z_n = - z_p$ we get $\phi_c = 2/3$
which would correspond to conditions of extremely high sample loading. If initial conditions 
are used such that $\phi > \phi_c$ in some regions, there will be an initial transient period that 
is not described by Eq~(\ref{eq4phi}), but once diffusion causes the analyte concentration
to drop below the critical  value, the subsequent evolution should follow Eq.~(\ref{eq4phi}). 

\section{Analysis of the one equation model}
\label{sec:analysis}
Eq.~(\ref{eq4phi}) in general describes a one dimensional nonlinear wave~\cite{whitam_book}. 
The nonlinearity arises because the local wave speed $V(\phi) = v/(1-\alpha \phi)$ changes 
with $\phi$. In particular, if $\alpha > 0$, larger values of $\phi$ correspond to higher 
speeds, so that if $\phi$ has a maximum at some point, then analyte 
ions near the maximum out runs the slower ions ahead of it causing the distribution of $\phi$ to lean 
forward and develop a sharp gradient at its leading edge (a shock wave). 
An analogous situation arises if $\alpha < 0$ except in this case the distribution of $\phi$ leans 
backwards as seen in Panels C and D of Figure~3. 
The diffusive term on the right hand side of Eq.~(\ref{eq4phi}) works to counter the sharpening 
of the wave due to the nonlinearity. When it is dominant the wave approaches a Gaussian shape 
in time with small departures from this limiting shape due to the nonlinearity. In case of small diffusivity 
the diffusive term is small everywhere except  in the vicinity of the shock; its presence  
prevents the sharp change in $\phi$ at the shock from evolving into a true mathematical discontinuity. 
In the remainder of this section we attempt to rationalize some of the observed facts 
about electromigration dispersion by exploiting some well known results from the mathematical theory 
of nonlinear waves.

\subsection{The weakly nonlinear case}
Let us suppose that $| \alpha \phi | \ll 1$ for all $x$ and $t$. Since $\phi$ is the ratio of analyte 
ion to that of the background ions, typically $| \phi | \ll 1$. The above condition can also be satisfied 
even when $|\phi|$ is not small compared to unity, provided $| \alpha |$ is small, that is, 
if the electrophoretic mobility of the 
analyte ion closely matches that of one of the background electrolytes. In either case, one may 
wish to neglect the term $ \alpha \phi$ in the denominator of Eq.~(\ref{eq4phi}), which would then describe
 waves that propagate with constant velocity $v$ while simultaneously undergoing diffusive spreading. This of course 
describes the usual ``linear'' electrophoresis. To see the effect of electromigration dispersion we 
must treat the term $ \alpha \phi$ as small, but not zero. This is achieved by making the approximation 
$(1 -  \alpha \phi)^{-1} \approx 1 +  \alpha \phi$ in Eq.~(\ref{eq4phi}), which then becomes: 
\begin{equation} 
\frac{\partial \phi }{\partial t} + v \frac{\partial \phi }{\partial x} + 2 \alpha v \phi  \frac{\partial \phi }{\partial x} 
= D \frac{\partial^{2} \phi}{\partial x^{2}}.
\label{burger_dimensional}
\end{equation} 
Eq.~(\ref{burger_dimensional}) takes a more familiar form if we describe it in a reference 
frame translating with velocity $v$, that is, we transform variables from $(x,t)$ to $(\xi,t)$ where 
$\xi = x - vt$. Eq.~(\ref{burger_dimensional}) then becomes the one dimensional Burgers' 
equation\footnote{in contrast to Eq.~(\ref{eq4phi}), the solutions of Burgers' equation are well 
behaved for all $\phi$, though the solution would correspond to physical reality only if $0 < \phi < \phi_{c}$}
\begin{equation} 
\frac{\partial \phi }{\partial t} +  2 \alpha v \phi  \frac{\partial \phi }{\partial \xi } 
= D \frac{\partial^{2} \phi}{\partial \xi^{2}}.
\label{burger}
\end{equation} 
Eq.~(\ref{burger}) is well known in fluid mechanics and is studied 
as a model for breaking water waves, one dimensional shocks in gas dynamics and the like. 
It  can be reduced~\cite{whitam_book} to the 
one dimensional diffusion equation through a nonlinear transformation of variables and thus solved exactly. 
The solution is: 
\begin{equation} 
\phi (\xi, t) = \left. \int_{-\infty}^{+\infty} \frac{\xi - \eta}{2 \alpha v t} \exp \{ - P(\eta; \xi, t) \} \, d \eta \right/
\int_{-\infty}^{+\infty} \exp \{- P(\eta; \xi, t) \} \, d \eta 
\label{cole-hopf_solution}
\end{equation} 
where 
\begin{equation} 
P(\eta; \xi, t) = \frac{(\xi - \eta)^{2}}{4 Dt} + \frac{\alpha v}{D} \int_{0}^{\eta} \phi (\xi^{\prime},0) \, d \xi^{\prime}.
\end{equation}
Given any initial profile for $\phi$ its subsequent time evolution may be calculated simply 
by evaluating the integrals  in Eq.~(\ref{cole-hopf_solution}). A special case of interest 
is 
\begin{equation}
\phi (x,0) = \Gamma \delta (x) 
\label{sharp_phi_ic}
\end{equation}
where $\Gamma = \int_{-\infty}^{+\infty} \phi(x,t) dx$ (a conserved quantity) and $\delta(x)$ is the 
Dirac delta function. This may be an appropriate model for $t \gg \tau_{d}$ when the peak width is 
much larger than its initial value so that the detailed shape  of the initial concentration distribution may 
be disregarded.  We will use the quantity $\Gamma$
to characterize the degree of sample loading. 
It has dimensions of length and may be attributed the following simple meaning: it is the 
length along the capillary that contains the same number of negative ions as there are 
analyte ions in the injected peak. 
Substitution of Eq.~(\ref{sharp_phi_ic}) in Eq.~(\ref{cole-hopf_solution}) yields
\begin{equation} 
\phi(x,t) = \frac{1}{2 \alpha v} \left( \frac{D}{\pi t} \right)^{1/2} F \left( \frac{x - vt}{\sqrt{4 Dt}} \right)
\label{solution_sharp_ic}
\end{equation} 
where 
\begin{equation} 
F(x) = \frac{A e^{-x^2}}{1 + (A/2) {\rm erfc} (x) }
\label{funcF}
\end{equation} 
with ${\rm erfc} (x) = (2/\sqrt{\pi}) \int_{x}^{\infty} \exp (- t^2 ) \, dt$ denoting the complementary error function. 
The constant $A$ is given by $A = e^{\alpha P} - 1$ where 
$P = \Gamma v / D$ is a  Peclet number 
based on the characteristic length scale $\Gamma$. If $\alpha P$ is small, $A \approx 0$ and in this case the 
denominator in Eq.~(\ref{funcF}) may be approximated by one. In this limit Eq.~(\ref{solution_sharp_ic}) 
reduces to a spreading Gaussian profile as expected. If $\alpha P$ is significant, $A$ is non-negligible 
and it has the same sign 
as $\alpha$. In this case the denominator in Eq.~(\ref{funcF}) creates an asymmetry in the profile, 
Eq.~(\ref{solution_sharp_ic}), causing it to steepen in the direction of peak propagation 
or create a steep trailing edge depending on the sign of $\alpha$. This behavior of course is well known from 
experimental and numerical studies of electromigration dispersion and is often referred to as ``fronting'' 
or ``tailing''.   It may be of interest to note that Eq.~(\ref{funcF}) is similar to a certain 
empirical function that has historically been used to fit experimental data on asymmetric 
peaks~\cite{erny_ac01,erny_JCA02}. 
However, unlike this so called ``Haarhoff-Van der Linde (HVL) function'' Eq.~(\ref{funcF}) 
is not an empirical fit but a mathematical consequence of the governing equations. 
The quantity $|\alpha |P$ that determines the strength of this effect may be expressed in terms of several 
relevant time scales: $|\alpha| P  = \tau_{e} \tau_{d} / \tau_{a}^{2}$ where $\tau_{a} = a/v$ is an 
advective time,  $\tau_{d}$ is the diffusive time scale introduced earlier and 
$\tau_{e} = |\alpha| \Gamma/ v$  is a new time scale associated with sample loading. The importance of the scale 
$\tau_{e}$ will become apparent later, but here we provide some numerical estimates. If we assume 
that the analyte peak is a Gaussian with a $10 \, \mu$m width and a peak concentration that is 
$2$ percent of the background, then $\tau_{e} \sim 0.5$ ms, if $v \sim 1$ mm/s and $|\alpha| \sim 1$.
The other two timescales using the estimates provided earlier are $\tau_{a} \sim 10$ ms and $\tau_{d} \sim 100$ ms.
Thus, in this example, $P \sim 0.5$ and therefore $A \sim 1$ which would correspond to ``high'' sample loading and on the basis of Eq.~(\ref{funcF}) 
strong deviations from Gaussianity may be expected. Thus, the quantity $|\alpha| P$ 
may be regarded as a suitable measure of sample loading in the context of peak distortion.

Another useful quantity that may be 
calculated~\cite{whitam_book} from the theory of nonlinear waves is the time needed 
to develop a shock:
\begin{equation}
t_{{\rm shock}} = - \left[ 2 \alpha v \left( \frac{\partial \phi (x,0)}{\partial x} \right)_{{\rm min}} \right]^{-1}
\end{equation}
where the suffix `min' indicates the minimum value of the function (which is clearly 
negative for a unimodal distribution). If the initial profile is a Gaussian of standard deviation 
$w_{0}$ then the minimum is easily evaluated and we get 
\begin{equation} 
t_{{\rm shock}} = \sqrt{\frac{\pi e}{2}}  \frac{w_{0}^{2}}{\alpha v \Gamma} \sim \frac{\tau_{a}^{2}}{\tau_{e}}.
\end{equation}
If we substitute our numerical estimates, we get $t_{{\rm shock}} \sim 0.2$ s which is very 
short compared to the total transit time across the capillary, typically  tens 
of minutes. 

Shock like solution profiles may be obtained~\cite{whitam_book} from  Eq.~(\ref{solution_sharp_ic}) on taking the 
asymptotic limit of $D \rightarrow 0$. If $\alpha >0$ we have leading edge shocks:
\begin{eqnarray}
\phi (x,t) &=& \frac{1}{2 \alpha} \left[ \frac{x}{x_{b}} - 1  \:\:\right] \quad {\rm for} \quad x_{f} \geq x \geq x_{b},\label{profile_burger}\\
x_{b} &=& vt,\\
x_{f}  &=& x_{b} \left[ 1 + 2 \sqrt{\frac{\tau_e}{t} } \right] \label{xf_burger}
\end{eqnarray}
and if $\alpha < 0$, we have trailing edge shocks: 
\begin{eqnarray}
\phi (x,t) &=& \frac{1}{2 |\alpha|} \left[ 1 - \frac{x}{x_{f}}  \:\:\right] \quad {\rm for} \quad x_{f} \geq  x \geq x_{b},\label{profile_burger1}\\
x_{f} &=& vt,\\
x_{b}  &=& x_{f} \left[ 1 - 2 \sqrt{\frac{\tau_e}{t} } \right]. \label{xf_burger1}
\end{eqnarray}
Here $x_f$ and $x_b$ are the locations of the ``front'' and ``back'' of the distribution $\phi(x,t)$ which 
vanishes outside the interval $[x_b,x_f]$.

The profile, Eq.~(\ref{solution_sharp_ic}), may be used to calculate some useful quantities 
such as the location of the centroid:
\begin{equation}
x_{c}(t) =\frac{\int_{-\infty}^{+\infty} x \phi (x,t) \; dx}{\int_{-\infty}^{+\infty}  \phi (x,t) \; dx}
\end{equation}
and the variance:
\begin{equation}
\sigma^{2}_{c}(t) =\frac{\int_{-\infty}^{+\infty} (x - x_c)^{2} \phi (x,t) \; dx}{\int_{-\infty}^{+\infty}  \phi (x,t) \; dx}.
\label{varc}
\end{equation}
The evaluation of the integrals is facilitated on transforming to the new variable 
$\eta = (x - vt)/ \sqrt{4 Dt}$. The second moment about the centroid ($x_c$), Eq~(\ref{varc}), 
is related in a simple way to $\sigma^{2}_{0}$ which is defined as in Eq~(\ref{varc}) but with $x_c$ 
replaced by $x_0 = vt$. This relation (sometimes referred to as the ``parallel axis theorem'' 
in mechanics): $\sigma_{0}^{2} = (x_0 - x_c)^{2} + \sigma_{c}^{2}$ easily follows from 
Eq~(\ref{varc}). Its use simplifies the calculation of the second moment, and we get 
after some algebra: 
\begin{equation}
x_{c} (t) = vt + \sqrt{4 Dt} \: \frac{F_1}{F_0},
\label{xc}
\end{equation} 
and 
\begin{equation}
\sigma^{2}_{c} (t) = 4 Dt \left( \frac{F_2}{F_0} - \frac{F_{1}^{2}}{F_{0}^{2}} \right),
\label{sigma_square_c}
\end{equation} 
where $F_{k} = \int_{-\infty}^{+\infty} \eta^{k} F (\eta) \; d \eta$ is the $k$ th moment 
of the function $F(\eta)$ defined by Eq.~(\ref{funcF}).  
In general the $F_k$ need to be determined numerically 
except for $k=0$, in which case $F_{0} = \alpha P \sqrt{\pi}$ which is merely a statement 
of the fact that the solution, Eq~(\ref{solution_sharp_ic}), satisfies  the normalization 
implied  by Eq~(\ref{sharp_phi_ic}). The speed of the centroid $V_{c} = dx_c/dt$ readily follows
from Eq~(\ref{xc}):
\begin{equation} 
V_c(t) = v \left[ 1 \pm  \frac{F_1}{ \sqrt{\pi} |\alpha|^{3/2} P^{3/2}} \sqrt{\frac{\tau_e}{t}} \right]
\sim v \left[ 1 \pm  \frac{2}{3} \sqrt{\frac{\tau_e}{t}} \right] \quad  ( {\rm if} \quad P \rightarrow \infty).
\label{centroid_V}
\end{equation} 
with the $+$ or $-$ sign depending on whether $\alpha$ is  positive or negative.
If the effective diffusivity is defined by $\sigma_{c}^{2} (t) = 2 D_{{\rm eff}} \: t$ then
\begin{equation}
D_{{\rm eff}} = 2 D \left( \frac{F_2}{F_0} - \frac{F_{1}^{2}}{F_{0}^{2}} \right) \sim   
\frac{1}{9} |\alpha| \Gamma v \quad  ( {\rm if} \quad P \rightarrow \infty).
\label{Deff}
\end{equation}
The asymptotic forms for large $P$ following the symbol $\sim$  in Eq.~(\ref{centroid_V})
and Eq.~(\ref{Deff}) are 
easily obtained on calculating 
the moments $F_k$ using the profiles Eq.~(\ref{profile_burger}) and Eq.~(\ref{profile_burger1}).
Thus, we see that the centroid velocity approaches $v$ asymptotically for times 
that are large compared to $\tau_{e}$, however the spread of the peak is diffusive 
at all times and may be described by the effective diffusivity indicated by Eq~(\ref{Deff}).
The phrase ``all times'' here refers to times that are nevertheless large compared to $\tau_d$ since 
otherwise Eq~(\ref{solution_sharp_ic}) may not be applied and the solution in this regime depends 
on the detailed features of the initial profile.

\subsection{The strongly nonlinear case}
The assumption $| \alpha \phi| \ll 1$ may not be a good one especially during the 
early part of the evolution in the case of high sample loading. The replacement of 
Eq.~(\ref{eq4phi}) by Burgers' equation, Eq.~(\ref{burger}), is not 
a good approximation in that case and we refer to this as the strongly nonlinear regime.
However,  an analytical solution is still possible if ${\rm Pe} \gg 1$. In this case, the diffusion term in 
Eq.~(\ref{eq4phi}) may be neglected  except in 
the immediate vicinity of a shock. Since ${\rm Pe}$ is usually large in electrophoresis 
it is reasonable to treat the problem as one of propagation of a shock front for times 
$t > t_{{\rm shock}}$, which formally corresponds to setting the right hand side 
of Eq.~(\ref{eq4phi}) to zero while admitting discontinuous functions within the class of 
solutions. It should be pointed out that Eq.~(\ref{eq4phi}), with $D=0$, and its solution Eq.~(\ref{profile}) had been noted earlier 
by others~\cite{mi_ev_ve_79a,math_th_elph_bk}, though its consequences for electromigration 
dispersion had not been adequately worked out. We do so here.

A solution to Eq.~(\ref{eq4phi}) with $D=0$ may be sought\footnote{the existence of such similarity 
solutions is suggested by the invariance of the equations under a scale transformation: $x \rightarrow x/\rho$
and  $t \rightarrow t/\rho$ for any nonzero constant $\rho$}
in similarity form in terms of the variable $X = x/vt$, that is, we look for solutions 
of the type $\phi(x,t) = \Phi (X)$ where $\Phi (X)$ is an unknown function. 
Substitution of this form in Eq.~(\ref{eq4phi}) with $D$ set to zero results in the following requirement
that $\Phi (X)$ needs to satisfy:
\begin{equation} 
\frac{d}{d X} \left[ \frac{\Phi}{1 - \alpha \Phi} \right] = X \frac{d \Phi}{d X},
\end{equation}
which may be easily rearranged as 
\begin{equation} 
 \left[ X - \frac{1}{(1 - \alpha \Phi)^{2}} \right] \frac{d \Phi}{d X} = 0.
\end{equation}
Thus, $\Phi$ must either satisfy the differential equation $ d \Phi / d X = 0$ and therefore 
must be a constant  or an algebraic 
equation that is easily solved: 
\begin{equation} 
\Phi (X) = \frac{1}{\alpha} \left[ 1 \pm \frac{1}{\sqrt{X}} \right].
\label{PhiX}
\end{equation} 
It is clear that neither of these functions by itself can represent the solution for 
all $X$ but rather each describes a piece of it.
The branch $\Phi$ is  constant must 
describe the solution outside the analyte zone and because 
of the requirement $\Phi( \pm \infty) = 0$ the constant must be zero. 
The solution within the analyte zone is best discussed by considering the 
cases corresponding to the two signs of $\alpha$ separately. \\[2ex]

\subsubsection{Leading edge shocks ($\alpha > 0$)}
The solution branch Eq.~(\ref{PhiX}) can be matched at one of its boundaries 
to the branch $\Phi(X)=0$ only if we accept the negative sign. This boundary 
then corresponds to $X=1$ which must be identified with the trailing edge. 
The leading edge is at some unknown position $X=X_{f} > 1$. We can determine 
$X_{f}$ by exploiting the condition that the integral of $\phi$ is a conserved 
quantity, which immediately follows on integrating Eq.~(\ref{eq4phi}):
\begin{equation}
\Gamma = \int_{-\infty}^{+\infty} \phi (x,t) \, dx = vt \int_{1}^{X_{f}} \Phi (X) \, dX 
\end{equation}
It is best to present the solution in terms of our original variables: 
\begin{eqnarray}
\phi (x,t) &=& \frac{1}{\alpha} \left[ 1 - \sqrt{\frac{x_{b}}{x}} \:\:\right] \quad {\rm for} \quad x_{f} \geq x \geq x_{b},\label{profile}\\
x_{b} &=& vt,\\
x_{f} &=& vt \left[ 1 + \sqrt{\frac{\tau_{e}}{t} } \:\: \right]^{2}  \label{xf}
\end{eqnarray}
and $\phi(x,t)=0$ outside the interval $[x_b,x_f]$. Here $\tau_{e} = |\alpha| \Gamma/ v$ is the time scale introduced earlier. 
Two quantities of particular importance in 
electrophoresis are the maximum value of $\phi$, $\phi_{m}(t) = \phi(x_{f},t)$ and 
the peak width $w = x_{f} - x_{b}$. These may be easily calculated from Eq.~(\ref{profile}) - (\ref{xf}): 
\begin{eqnarray} 
\phi_{m} (t) &=& \frac{1}{ \alpha} \left[ 1 + \sqrt{ \frac{t}{\tau_{e}} } \: \: \right]^{-1} \label{peak_value}\\
w(t) &=& v \sqrt{\tau_{e} t} \left[ 2 + \sqrt{\frac{\tau_{e}}{t} } \:\: \right] \label{peak_w+}
\end{eqnarray}
The propagation speed of the peak is 
\begin{equation} 
V_{f} = \frac{d x_f}{dt} = v \left[ 1 + \sqrt{\frac{\tau_{e}}{t} } \:\: \right].
\label{shock_speed}
\end{equation}
It is easy to verify by direct substitution that Eq.~(\ref{peak_value}) and Eq.~(\ref{shock_speed}) 
satisfy the Rankine-Hugoniot condition for shock waves~\cite{whitam_book}. 

It is clear from these formulas that in the asymptotic limit of long  times ($t \gg \tau_{e}$)
the wave propagates unchanged in shape with its width increasing in proportion to the square root of time and its 
amplitude decreasing in inverse proportion to the square root of time while its integral is conserved. This is classical 
diffusive behavior, and therefore an effective diffusivity ($D_{{\rm eff}}$) 
can be introduced through the asymptotic relation $\sigma^{2} \sim 2 D_{{\rm eff}} \: t$ for the increase of variance
with time. To determine 
$D_{{\rm eff}}$ one needs to calculate the variance of the distribution represented by Eq.~(\ref{profile}) 
and then determine the leading order term in its asymptotic expansion in the small parameter $\tau_{e}/t$. This may be done 
and leads after some algebra to the rather simple final result
\begin{equation} 
D_{{\rm eff}} = \frac{1}{9} |\alpha| \Gamma v.
\label{Deff_shock}
\end{equation} 
We write $|\alpha|$ in place of $\alpha$ in anticipation of the fact that this expression is equally 
valid for a trailing edge shock ($\alpha < 0$). This is seen to be identical to the large Peclet number ($P$) 
limit of the effective diffusivity obtained from the solution of Burgers' equation indicated in Eq.~(\ref{Deff}).

\subsubsection{Trailing edge shocks ($\alpha < 0$)}
The requirement that $\Phi$ must be non-negative implies that only the 
negative sign in Eq.~(\ref{PhiX}) corresponds to physically realizable solutions. Thus,
\begin{equation} 
\Phi (X) = \frac{1}{|\alpha|} \left[ \frac{1}{\sqrt{X}} - 1 \right]
\label{PhiX1}
\end{equation} 
where $X \leq 1$. Therefore in this case, $X=1$ actually corresponds to the leading 
edge and $X=X_{b} < 1$ is the trailing edge which is where the shock is located. 
To determine $X_b$ we once again use the constancy of the integral of $\phi$:
\begin{equation}
\Gamma = \int_{-\infty}^{+\infty} \phi (x,t) \, dx = vt \int_{X_{b}}^{1} \Phi (X) \, dX.
\end{equation}
The relations analogous to Eq.~(\ref{profile}) - (\ref{xf}) are
\begin{eqnarray}
\phi (x,t) &=& \frac{1}{|\alpha|} \left[ \sqrt{\frac{x_{f}}{x}}  - 1 \:\:\right] \quad {\rm for} \quad x_{f} \geq x \geq x_{b},\label{profile1}\\
x_{b} &=& vt \left[ 1 - \sqrt{\frac{\tau_{e}}{t} } \:\: \right]^{2} \label{xb1}\\
x_{f} &=& vt. \label{xf1}
\end{eqnarray}
The amplitude, width and  speed of the peak may be calculated as before, they are
\begin{eqnarray} 
\phi_{m} (t) &=& \frac{1}{ |\alpha|} \left[ \sqrt{ \frac{t}{\tau_{e}} } - 1 \: \: \right]^{-1} \label{peak_value1}\\
w(t) &=& v \sqrt{\tau_{e} t} \left[ 2 -  \sqrt{\frac{\tau_{e}}{t} } \:\: \right]  \label{peak_w-}\\
V_{b} &=& \frac{d x_b}{dt} = v \left[ 1 - \sqrt{\frac{\tau_{e}}{t} } \:\: \right].
\label{shock_speed1}
\end{eqnarray}
The effective diffusivity $D_{{\rm eff}}$ is once again given by Eq.~(\ref{Deff_shock}).

Since Burgers' equation follows from Eq.~(\ref{eq4phi}) if $\phi$ is small, one might expect that 
Eq.~(\ref{profile1}) - (\ref{xf1}) should reduce to Eq.~(\ref{profile_burger1}) - (\ref{xf_burger1})
at asymptotically large times. This is indeed true. To show this, first observe that from Eq.~(\ref{xb1})
\begin{equation}
x_{b} = vt \left[ 1 - \sqrt{\frac{\tau_{e}}{t} } \:\: \right]^{2} = vt \left[ 1 - 2 \sqrt{\frac{\tau_{e}}{t} } + \frac{\tau_{e}}{t} \right]
\sim vt \left[ 1 - 2 \sqrt{\frac{\tau_{e}}{t} }  \right]
\end{equation}
at large times.
Secondly, if we put $x=x_f - \xi$ and regard $\xi/x_f$ as small (since $w \sim \sqrt{t}$ while $x_f \sim t$), then
\begin{equation} 
\phi (x,t) = \frac{1}{|\alpha|} \left[ \sqrt{\frac{x_{f}}{x}}  - 1 \:\:\right]
= \frac{1}{|\alpha|} \left[ \left( 1 - \frac{\xi}{x_f} \right)^{-1/2}   - 1 \:\:\right]
\sim \frac{\xi}{2 |\alpha| x_f } = \frac{1}{2 |\alpha|} \left( 1 - \frac{x}{x_f} \right).
\end{equation}
Thus, Eq.~(\ref{profile_burger1}) - (\ref{xf_burger1}) are recovered. Similarly, if $\alpha > 0$,
Eq.~(\ref{profile}) - (\ref{xf}) go over to  Eq.~(\ref{profile_burger}) - (\ref{xf_burger}) at large times.

\section{Numerical simulations}
\label{sec:simulations}
In order to test the accuracy of some of the results presented in the previous section, they 
were compared to solutions of Eq.~(\ref{eq4phi}) computed numerically using 
a second order (central) finite difference algorithm for spatial derivatives coupled with a fourth order 
Runge-Kutta time stepping scheme with fixed grid and step size. We will refer to this 
numerical calculation as ``the full solution''.

As a test problem, an initial profile $\phi(x,0)= \phi_{m} \exp(-x^{2}/2 \sigma_{0}^{2})$ was specified and 
$\phi_{m}$ was varied keeping $\sigma_{0}$ fixed. This involves no loss of generality because if Eq.~(\ref{eq4phi})
is rewritten in terms of dimensionless space and time variables: $x_{*} = x / \sigma_{0}$ 
and $t_{*} = v t / \sigma_{0}$, then it is easily seen that the only parameters that enter into the 
problem are: $\alpha$, $P = v \Gamma / D$ and $\Gamma / \sigma_{0} = \sqrt{2 \pi}  \, \phi_{m}$.
Comparisons were made over a wide range of parameter values, but for brevity, we present here only 
a few cases that are illustrative of the general trend.

\begin{figure*}
  \includegraphics[width=1.0\textwidth]{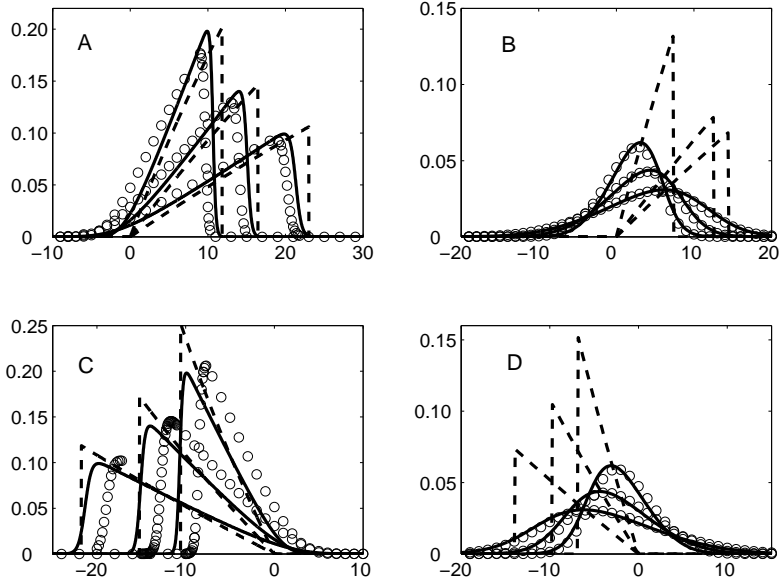}
\caption{Comparison of the normalized concentration  $\phi$ (y-axis) as a function of $(x - vt)/\sigma_{0}$ (x-axis) 
obtained from the full solution, Eq.~(\ref{eq4phi}) (open circles), Eq.~(\ref{solution_sharp_ic}) (solid lines) 
and Eq.~(\ref{profile}) or Eq.~(\ref{profile1}) (dashed lines) at fixed instants of time 
$v t / \sigma_{0} = 50, 100$ and $200$. The parameters are: 
$P=62.7$: $\alpha=0.5$ (Panel A), $\alpha=-0.5$ (Panel C) and  $P=5.0$: 
$\alpha=0.5$ (Panel B), $\alpha=-0.5$ (Panel D).}
\label{fig:2}       
\end{figure*}
In Figure~\ref{fig:2} we compare the analyte concentration profiles obtained by numerically evolving Eq~(\ref{eq4phi}) with 
the analytical formulas Eq.~(\ref{solution_sharp_ic}) and Eq.~(\ref{profile}) or Eq.~(\ref{profile1}).
 It is seen that for large values of $P$ all of the profiles are very close.
For smaller value of $P$, the full solution is in good agreement with the analytical solution 
of the Burgers' equation, Eq.~(\ref{solution_sharp_ic}), but does not agree with the shock profiles 
Eq.~(\ref{profile}) or Eq.~(\ref{profile1}). This is because the dynamics is dominated by diffusion.

\begin{figure*}
  \includegraphics[width=1.0\textwidth]{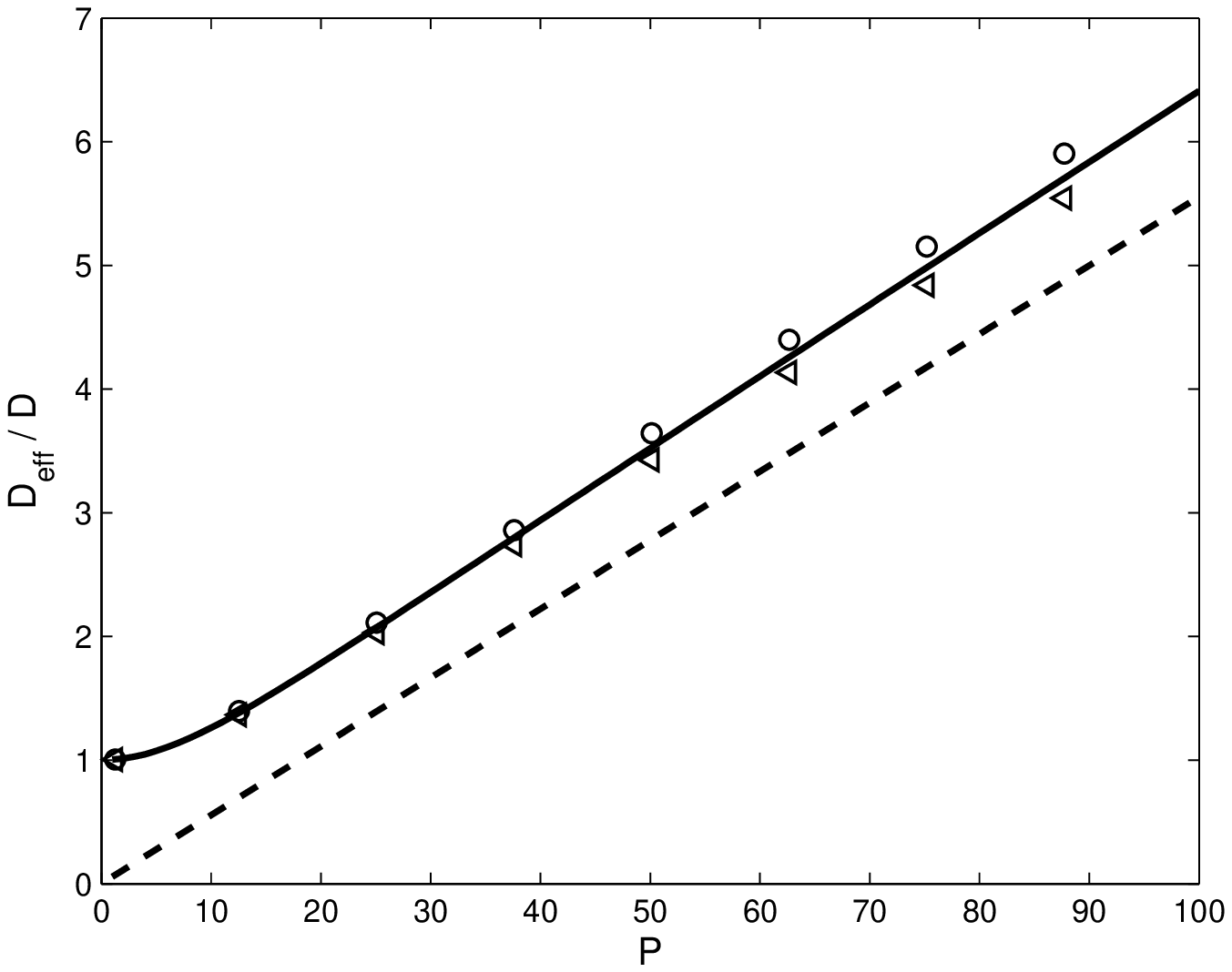}
\caption{The normalized effective diffusivity  as a function of the Peclet number $P$. The solid line 
is from Eq.~(\ref{Deff}) and the dashed line is from Eq.~(\ref{Deff_shock}). The symbols indicate values 
calculated from the numerical solution of Eq.~(\ref{eq4phi}). Two cases are shown, $\alpha = + 0.5$ (circles)
and $\alpha = - 0.5$ (triangles).}
\label{fig:3}     
\end{figure*}
The formulas given by Eq.~(\ref{Deff}) and Eq.~(\ref{Deff_shock}) could be of 
some practical value so we compared them with the corresponding results from the 
full solution in Figure~\ref{fig:3}. To do this, the variance $\sigma^{2}$ was calculated from the concentration 
profile in the full solution and the rate of change of variance $d (\sigma^{2})/dt$ was plotted as a function of 
time $t$. In all cases, this quantity was found to asymptote to a constant value, and equating 
this value to $2 D_{{\rm eff}}$ gave an effective value of the diffusivity. These are shown in 
Figure~\ref{fig:3} by the symbols for a range of values of the Peclet number $P$, for $\alpha = \pm 0.5$.
It should be observed that $D_{{\rm eff}}$ is independent of the sign of $\alpha$ as predicted 
by the theoretical results. The small difference in the value of $D_{{\rm eff}}$ for positive and 
negative $\alpha$ is due to the fact that the peak width approaches its asymptotic value 
from above or from below depending on the sign of $\alpha$ as is suggested by 
Eq.~(\ref{peak_w+}) and Eq.~(\ref{peak_w-}) for the peak width.  Thus, the numerical
determination based on the full solution results in a slight over estimate of 
the asymptotic value in one case and a slight under estimate in the other. The numerical calculation is seen to 
be in good agreement with Eq.~(\ref{Deff}). However, the dashed line has a constant offset with respect to the 
solid line because when $P$  is large, $D_{{\rm eff}}/D =1+\frac{1}{9}|\alpha| P + \cdots$,
and the limit represented by Eq.~(\ref{Deff_shock}) is obtained by neglecting the first term 
in relation to the second.  It is therefore a valid approximation and has a low relative error when 
$P$ is much larger than unity, but if this is not the case then one must use Eq.~(\ref{Deff}).

\section{Concluding remarks}
In this paper a simple model is constructed and rigorously solved with the objective 
of providing a physical understanding of the essential features observed in the CE 
signal at high sample loading. The model leaves out many features that are undoubtedly 
present in any real experiment, such as, the presence of many more than three ionic species,
shifts in ionic equilibria, differential diffusion between species and the like.  On the other 
hand, the fact that the principal qualitative features of the signal may be reproduced in this highly 
simplified model indicates that the neglected effects, though present, are not causative 
but only incidental to the problem.

The main results that follow from this analysis 
that might have some validity in a broader context are:
\begin{enumerate} 
\item Leading edge shocks are to be expected when the analyte ion has an
electrophoretic mobility that is intermediate to that of the co and counter ions and a trailing edge shock 
should be seen otherwise.
\item A peak showing a leading edge shock will elute earlier and a peak showing a trailing edge shock will
elute later compared to the 
arrival times in an otherwise identical experiment where sample loading is kept low.
\item The quantity $|\alpha| P$, where $P = \Gamma v / D$ is a Peclet number
based on the length scale $\Gamma$, completely determines the nature of the wave. 
If $|\alpha| P  << 1$ the peak would evolve into a Gaussian with a small amount of asymmetry and 
if $|\alpha| P >> 1$ a shock like structure is formed. It is the single controlling parameter 
that must be kept low in order to mitigate the effects of electromigration dispersion. 
\item At long times the peak spreads diffusively with an effective diffusivity $D_{{\rm eff}}$. 
The ratio $D_{{\rm eff}}/D$ depends on the Peclet number $P$; it is proportional to $P$ when $P$ is large. 
\item The peak asymmetry that is characterized by the quantity $A$ has an exponential 
dependence on the Peclet number, $P$: $A = e^{\alpha P} - 1$, as a result, the transition 
from a ``weak asymmetry'' to a ``strong shock'' as one increases the sample loading is very 
sharp.
\item If the analyte electrophoretic mobility is closely matched to the electrophoretic mobility 
of one of the background species (this makes $|\alpha|$ small) peak distortion and dispersion 
can be minimized.
\end{enumerate} 
These observations appear to be consistent with laboratory experiments~\cite{bouskova_elph04,horka_elph00} and results from 
numerical simulations~\cite{mikkers_ac99}. The analytical formulas presented in this paper can be checked quantitatively through 
carefully designed experiments that closely mimic the simplifying assumptions made in this study.
It is hoped that this paper will motivate efforts in that direction.
\begin{acknowledgements}
This manuscript was prepared when the first author was on sabbatical at the Institute for Mathematics and its Applications (IMA) 
at the University of Minnesota and thanks the IMA for their kind hospitality. We also thank Professor 
Joseph B. Keller for reading and commenting on a draft of the manuscript.
\end{acknowledgements}

\bibliographystyle{spmpsci}       

%
%

\end{document}